# Ordered spin-ice state in the geometrically frustrated metallic-ferromagnet $Sm_2Mo_2O_7$

Surjeet Singh[1*], R. Suryanarayanan[1], R. Tackett[2], G. Lawes[2], A. K. Sood[3], P. Berthet[1], A. Revcolevschi[1]

[1]Université Paris Sud, Laboratoire de Physico-Chimie de l'Etat Solide, UMR8182, Bât 414, Orsay, France.
[2]Department of Physics and Astronomy, Wayne State University, Detroit, MI 48202 USA.
[3]Department of Physics, Indian Institute of Science, Bangalore 560012, India.

The recent discovery of Spin-ice is a spectacular example of non-coplanar spin arrangements that can arise in the pyrochlore $A_2B_2O_7$ structure. We present magnetic and thermodynamic studies on the metallic-ferromagnet pyrochlore $Sm_2Mo_2O_7$. Our studies, carried out on oriented crystals, suggest that the Sm spins have an ordered spin-ice ground state below about $T^* = 15$ K. The temperature- and field-evolution of the ordered spin-ice state are governed by an antiferromagnetic coupling between the Sm and Mo spins. We propose that as a consequence of a robust feature of this coupling, the tetrahedra aligned with the external field adopt a "1-in, 3-out" spin structure as opposed to "3-in, 1-out" in dipolar spin ices, as the field exceeds a critical value.

**PACS: 75.30 –m, 75.30.Kz, 75.40Cx, 81.30-h**

Geometrical frustration arises when the magnetic interactions on a lattice are incompatible with the underlying crystalline symmetry. Systems with magnetic ions on triangular or tetrahedral lattices can exhibit geometrical frustration, which is often relieved at sufficiently low temperatures by weak residual interactions giving way to complex magnetic ground states[1]. One particularly important example of geometrical frustration arises in materials based on the pyrochlore structure ($A_2B_2O_7$), where A and B ions form networks of corner-linked tetrahedra displaced by half the unit cell dimensions from each other. Among the several interesting results recently reported in these materials[2], one of the most remarkable is the discovery of spin-ice state in insulating pyrochlores $Dy_2Ti_2O_7$ (DTO) and $Ho_2Ti_2O_7$ (HTO)[3-5], having a disordered ground state in which two spins on each tetrahedron point-in and two spins point-out, isomorphic to proton ordering in water-ice[6].

The recent observation of the spin ice state in the metallic-ferromagnet pyrochlore $Nd_2Mo_2O_7$ (NMO) has opened-up an entirely new set of avenues for the future research in frustrated systems with itinerant degrees of freedom[7-9]. The spin-chirality in NMO is believed to perturb the charge dynamics so strongly that unconventional transport properties emerge in the spin-ice state. The gigantic anomalous Hall component which appears in the spin-ice state in NMO, for example, cannot be described using any of the conventional theories of anomalous Hall effect[7, 10]. While the spin-chirality mechanism for probing charge transport in the presence of complex spin-textures is an important development, the lack of similar systems that couple the itinerant and the magnetic degrees of freedom on a frustrated lattice hampered further research in this field. In this Letter we have attempted to fill this void by presenting a second example of a spin-ice state in the metallic pyrochlore $Sm_2Mo_2O_7$ (SMO).

SMO is a metallic-ferromagnet pyrochlore with a ferromagnetic temperature ($T_C$) of 80 K due to Mo spins ordering[11]. However, the ground state properties of the Sm spins in SMO remained unexplored due to the large neutron absorption cross-section of Sm, which precludes neutron scattering studies and the difficulties in growing large single-crystals. Our investigations on oriented crystals of SMO show that the Sm spins in this compound have an ordered spin-ice ground state that develops below 15 K. The spin-ice state in SMO results from a robust antiferromagnetic (AF) coupling of the Sm spins with Mo spins. Our observations suggest that the strength of AF f-d coupling between the rare-earth and Mo spins in SMO is much stronger than in NMO, which dramatically affects the evolution of the ordered spin-ice state under an applied magnetic field.

The single-crystal of SMO was grown using the optical floating-zone method in a purified Ar atmosphere. The successful growth of cm$^3$ size crystals of SMO is achieved by overcoming specific difficulties including the decomposition of the pyrochlore phase at low temperatures[12], the highly volatile nature of $MoO_2$, and the dependency of the oxidation state of Mo on small variations in the growth atmosphere. The magnetization measurements were done with Quantum Design SQUID magnetometer. The specific heat (C) was measured using a standard quasi-adiabatic technique on a Quantum Design PPMS. The lattice contribution ($C_{latt}$) to the C of SMO is approximated by the specific heat of non-magnetic analogue $La_2Zr_2O_7$ after molar mass correction.

The temperature dependence of the magnetization M(T) of SMO under several magnetic fields applied parallel to the [111] direction is shown in Fig. 1. A step-like increase in M(T) below 80 K marks the onset of FM ordering in the Mo sublattice. Below $T_C$, the magnetization exhibits two prominent features: a broad maximum centered around 35 K, observed only at low fields, followed by a sharp drop near 15 K. We argue that the low temperature features in M(T) of SMO arise from AF correlations developing between the Sm and Mo spins. The AF coupling between the Sm and Mo spins can be inferred from the M(H) isotherms plotted in Fig. 1. The magnitude of the T = 5 K magnetization is dramatically suppressed compared to the T = 30 K value for a field applied along the [111] direction. This likely arises from the tendency of Sm spins to align antiferromagnetically to the ferromagnetically ordered Mo spins.

In the pyrochlore SMO, the trigonal symmetry ($D_{3d}$) of the CEF splits the $^6H_{5/2}$ ground state in a free $Sm^{3+}$ ion into three Kramer's doublets. From Raman scattering and specific heat studies we estimated that the excited CEF levels in the analogous titanate ($Sm_2Ti_2O_7$) are located well above the ground state doublet at T = 90 K and 175 K[13]. As the anionic arrangement around Sm is similar in both the compounds, we expect the Kramers doublet ground state of Sm in SMO to be an anisotropic well-isolated doublet.

The anisotropic magnetic behavior is evident from Fig. 2 where M(H) isotherms are shown for magnetic fields along the [100], [110], and [111] axes at T = 2 and 25 K. While the magnetization behavior is nearly isotropic at 25 K, highly anisotropic behaviors emerge at 2 K with easy- and hard-axis behaviors along [100] and [110], respectively, and a spin-flip transition in the [111] magnetization at $H_C \sim 15$ kOe. Since the FM state due to Mo spin ordering is nearly isotropic, as evidenced by the 25 K magnetization and the temperature variation of M(T) above T* = 15 K along the [100] and [111] directions (inset, Fig. 2d), the huge anisotropy is likely due to the CEF induced single-ion properties of the Sm spins ordering below T*. In the pyrochlore structure, due to a severe trigonal distortion along the local [111] axes, the axial component of the CEF tends to confine the spins either along the local [111] axes of an elementary tetrahedron (Ising anisotropy) or in planes perpendicular to the [111] directions (planar anisotropy). The Dy(Nd) spins in DTO(NMO), for example, are Ising-like[7, 8].

We note that the nature of magnetic anisotropy in SMO below T* (Fig. 2) bear a striking similarity to the dipolar spin-ices DTO[14] and HTO[15], and the spin-ice ground state of Ising FM model on the pyrochlore lattice due to Harris et al.[16]. This suggests that the Sm spins in the pyrochlore SMO are Ising-like having a spin-ice like ground state. However, the temperature T* below which the spin-ice like correlations develop in SMO is much higher than the corresponding temperature in dipolar spin-ices (T* ~ 1 K). This is because an effective FM force between the Sm spins results from their AF coupling with the ordered Mo spins. Further, the "2-in, 2-out" configurations on each Sm tetrahedron in SMO are expected to be identical in the presence of uniform internal magnetic field of Mo sublattice, akin to ordered "2-in, 2-out" configuration realized in dipolar spin-ices under externally applied magnetic fields[17].

The spin-ices under magnetic fields (i.e. in their ordered spin-ice state) show remarkable irreversibility in magnetization behavior. Hysteresis loops in DTO are found to be reversible above 0.65 K but highly irreversible at lower temperatures with coercivity of 2 kOe[18, 19]. The ordered spin-ice state in SMO exhibits a similar irreversibility (Fig. 3) with the coercive field rising sharply to a value of 2.5 kOe below T* = 15 K (inset, Fig. 3).

Fig.4(inset) shows the temperature variation of C/T and $C_{mag}$/T in SMO. The anomaly at $T_C$ = 78 K arises from the FM ordering of the Mo spins. A broad Schottky anomaly is found at lower temperatures as the systems enters the ordered spin-ice state similar to what is observed in NMO. We argue that the "2-in, 2out" ordering in these compounds is essentially a non-cooperative phenomenon that occurs over a broad temperature range. This is reflected in the neutron scattering studies where the intensity of the magnetic Bragg peaks in NMO due to $Nd^{3+}$ moments ordering in a "2-in, 2-out" arrangement began to grow slowly near 40 K and continue to grow monotonically down to at least 8 K[7].

The change in magnetic entropy in SMO in the [2, T] K range, $\Delta S_{mag}$(T) (Fig. 4), obtained by integrating $C_{mag}$/T, reaches 87 % of Rln2 at T*. Since, in the temperature range T < T* the $C_{mag}$ is mainly due to the Sm spins[20], we recover nearly the full Rln2 entropy associated with the Kramers doublet ground state. In dipolar spin-ices, where the "2-in, 2-out" ground state is highly degenerate, the spin entropy in zero-field saturates to a value that is nearly 30 % short of Rln2, consistent with the "missing entropy" predicted for water-ice[5]. The absence of missing entropy in SMO, as also in dipolar spin-ices under applied fields[21], is consistent with an ordered "2-in, 2-out" ground state for SMO. The increase in $\Delta S_{mag}$ above Rln2 at higher temperatures is presumably due to additional contributions to $C_{mag}$, including excited CEF levels of Sm and the Mo spins below $T_C$.

Under an applied field of 50 kOe along the [111] direction, the Schottky peak shifts to slightly lower temperatures. This small negative shift suggests that even at 50 kOe, the Zeeman splitting in SMO is determined mainly by the AF f-d exchange. In NMO the internal field changes sign from negative to positive when the external field exceeds 30 kOe[22]. This suggests that the AF f-d coupling in SMO is much stronger then in NMO which explains why the low-temperature peak in our specific heat studies and similar studies reported in ref[23] on polycrystalline sample up to 130 kOe shifts monotonically to lower temperatures. Finally, we note that the low temperature integrated entropy in SMO changes only very slightly between H = 0 and H =50 kOe (Fig. 4), consistent with the proposal that the H = 0 ground state of SMO is an ordered spin-ice state.

We shall now briefly discuss the saturation magnetization in SMO along the three field directions. In SMO, the expected saturation moment in the high-field limit along the three field directions is given by[24]: $\sigma_{100} = (\mu_{Mo} + 1/\sqrt{3}\ \mu_{Sm})$, $\sigma_{111} = (\mu_{Mo} + 1/2\ \mu_{Sm})$, $\sigma_{110} = (\mu_{Mo} + 1/\sqrt{6}\ \mu_{Sm})$, where $\mu_{Mo}$ and $\mu_{Sm}$ are the ordered Mo and Sm moment, respectively. In Fig. 2d, M(H) in SMO along the three field directions is shown on an enlarged scale. While $M_H$ along [100] saturates at $1.4\mu_B/(\frac{1}{2}\text{f.u.})$, the magnetization along [110] continues to increase with increasing field, exceeding the $1.4\mu_B/(\frac{1}{2}\text{f.u.})$ value above 50 kOe, contrary to expectations. This argues that the saturation value attained along [100] does not reflect the complete saturation magnetization. These results can be understood if we suppose that the AF f-d exchange forces the magnetization vectors on the two sublattices to remain antiparallel up to the highest fields. In this framework, $\mu_{Mo}$ and $\mu_{Sm}$ are calculated to be 1.54 and 0.25 $\mu_B$, respectively, slightly smaller than the free ion moments (2 and 0.7 $\mu_B$). Similarly reduced values for rare-earth and Mo moments are previously reported for $Nd_2Mo_2O_7$[25] and $(Tb_{1-x}La_x)_2Mo_2O_7$[26].

The spin-flip transition in the [111] magnetization of SMO can be understood in terms of a reorientation of the Mo spins from the [100] direction to the [111] direction above a critical field of $H_C$ =15 kOe. The robust f-d coupling produces a direction reversal of one of the Sm spins in the basal plane, leading to a "1-in, 3-out" spin-ordered state. This is illustrated schematically in Fig. 3b. It should be compared to the spin-flip transition under a [111] field in dipolar spin-ices, where the spin arrangement changes from "2-in, 2-out" to "3-in, 1-out" at the spin-flip transition, when the weak dipolar forces are overwhelmed by the increasing Zeeman energy.

To summarize, using bulk measurements on oriented crystals we argue that the Sm spins in the metallic-ferromagnet pyrochlore SMO have an ordered spin-ice ground state. The ordered spin-ice in SMO is non-dipolar in origin, but rather develops as a result of AF coupling of the Sm spins with ferromagnetically ordered Mo spins. This coupling is shown to be exceptionally robust. As a consequence of the robust feature of this coupling, the spin-flip transition in SMO under a [111] field is interpreted as a change in spin arrangement from "2-in, 2-out" to "1-in, 3-out" as opposed to "3-in, 1-out" in dipolar spin-ices. Because of the large effective FM interactions between the Sm ions, SMO offers an opportunity to investigate spin-ice dynamics at much higher temperatures than the dipolar systems. Furthermore, we propose SMO, with its spin-ice like ground state that shows a distinct spin-flip transition under [111] field, as an excellent candidate for testing the validity of the spin chirality mechanism of the Anomalous Hall effect.

**Acknowledgements**

We would like to thank Guy Dhalenne, Romuald Saint Martin, Jacques Berthon and Sudesh Dhar, for useful discussions. A major portion of this work is funded by the Centre Franco-Indien pour la Promtion de la Recherche Avancée (CEFIPRA) under project no. 3108-1. AKS thanks the Department of Science and Technology, India, for financial support. RT and GL thank the Jane and Frank Warchol Foundation and the Institute for Manufacturing Research at Wayne State University for financial support.

**Figure captions**

**Figure 1** (Color online) Temperature and magnetic field (inset) dependence of magnetization under several applied fields along [111] in a $Sm_2Mo_2O_7$ crystal.

**Figure 2** (Color online) Isothermal magnetization in a single crystal of $Sm_2Mo_2O_7$ at T = 2 and 25 K. Applied field along: **a,** [100]; **b,** [110]; and **c,** [111] crystallographic directions. **d.** The expanded view shows the magnetization along the three axes close to saturation. The inset in **d** shows temperature dependence of magnetization under a 10 kOe field applied along the [100] and [111] axes. **e.** The Schematics of the proposed mechanism for the spin-flip transition under [111] field in $Sm_2Mo_2O_7$. The spin configuration on the Sm tetrahedron changes from "2-in, 2-out" to "1-in, 3-out" as the applied fied (H) exceeds a critical value $H_C$. The resultant magnetic moment per Mo(Sm) tetrahedron is indicated by thick arrow at the centre of each tetrahedron.

**Figure 3** (Color online) Magnetic hysteresis loops in the [100] oriented crystals of $Sm_2Mo_2O_7$ at several temperatures between 2 K (outermost loop) and 75 K. Inset shows a sharp increase in the coercive field as determined from the main panel in the ordered spin-ice state below T*.

**Figure 4** (Color online) Specific heat (C), magnetic specific heat ($C_{mag}$) and change in magnetic entropy ($\Delta S_{mag}$) in $Sm_2Mo_2O_7$.

Fig. 1 (Singh et al.)

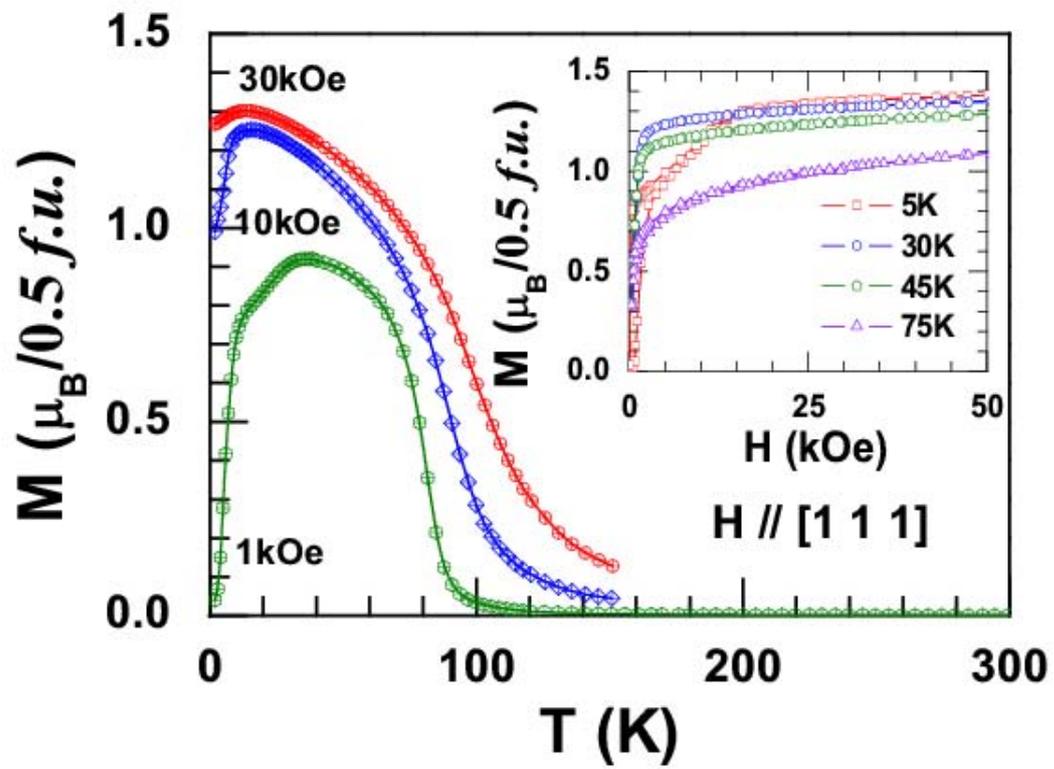

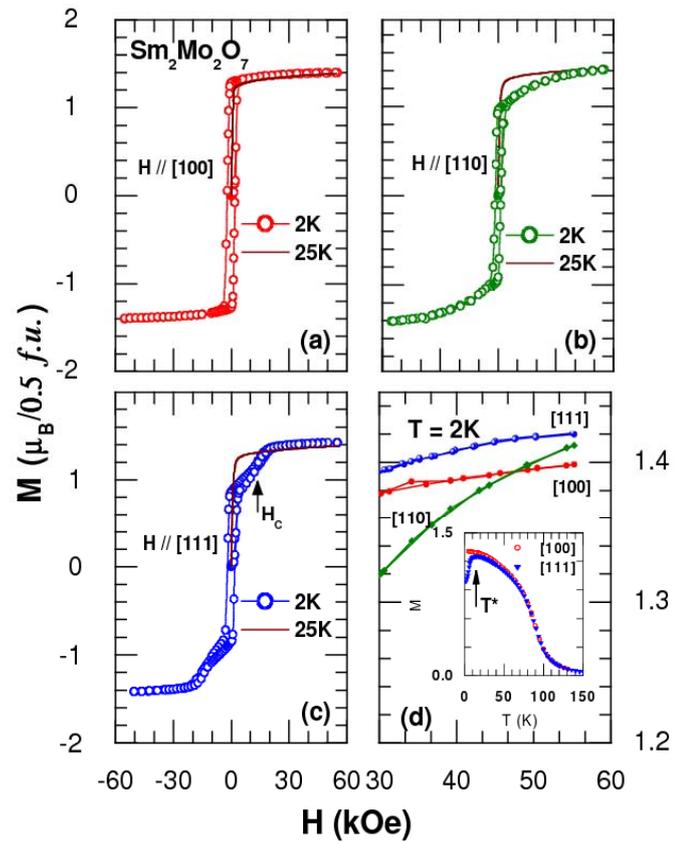

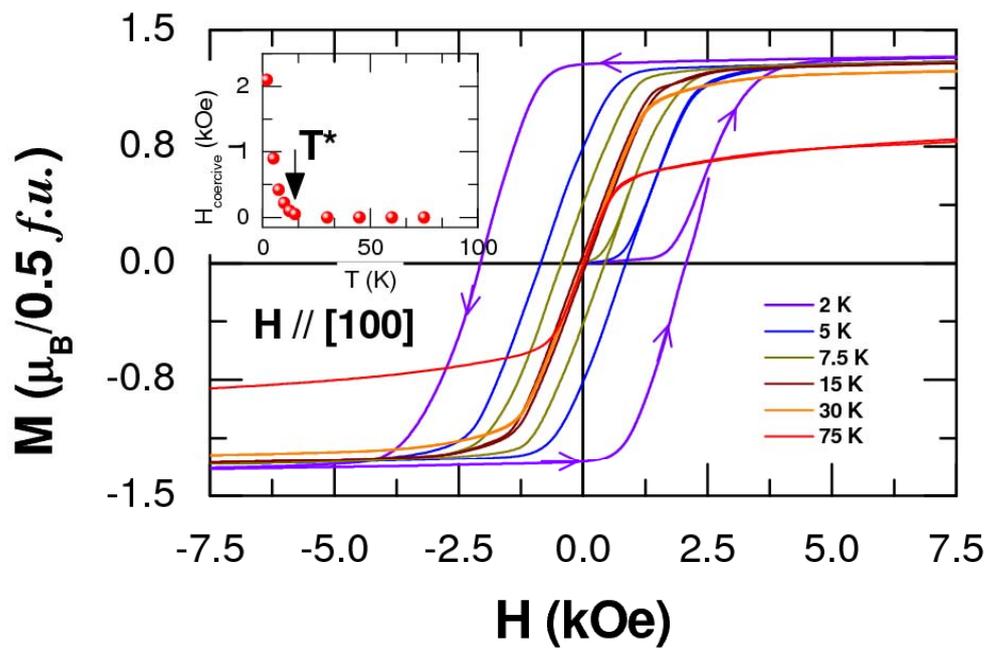

Fig. 4 (Singh et al.)

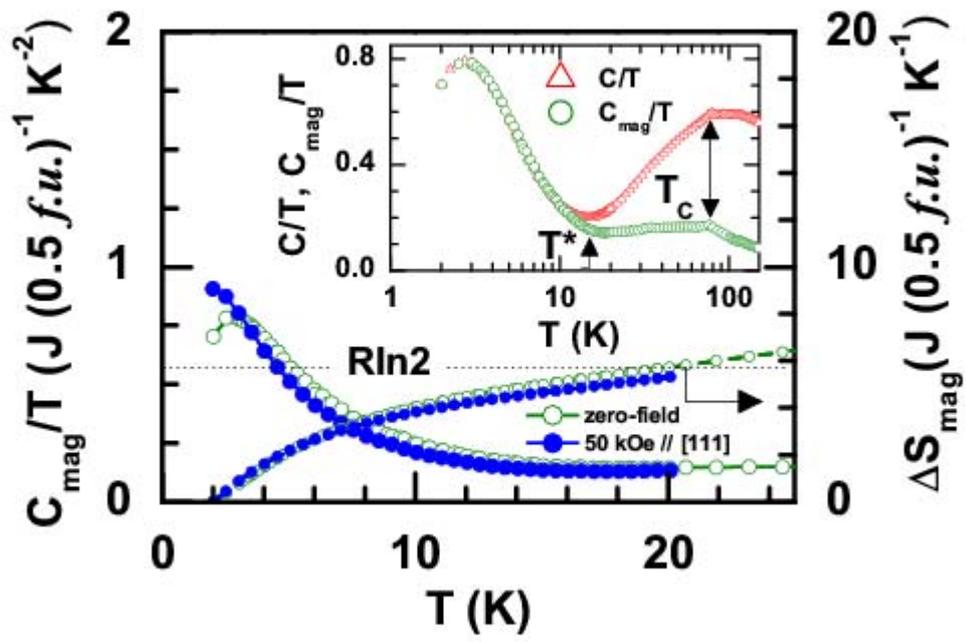